# Large scale clustering from Non-Gaussian texture models

E. Gaztañaga[1,3] & P. Mähönen[2,3]

[1] *CSIC, Centre d'Estudis Avançats de Blanes, c/ camí St. Barbara s/n, 17300 Blanes, Girona, SPAIN*
[2] *Department of Physical Sciences, Theoretical Physics, University of Oulu, 90570 Oulu, FINLAND*
[3] *Oxford University, Astrophysics, Keble Road, Oxford OX1 3RH, UK*

**ABSTRACT**

A basic ingredient to understand large scale fluctuations in our present day Universe is the initial conditions. Using N-body simulations, we compare the clustering that arises from Gaussian and non-Gaussian initial conditions. The latter is motivated by a global texture model which has initial $J$-order correlations $\overline{\xi}_J$ close to the strongly non-Gaussian regime $\overline{\xi}_J \simeq \overline{\xi}_2^{J/2}$. The final amplitudes $S_J \equiv \overline{\xi}_J/\overline{\xi}_2^{J-1}$ in the non-Gaussian (texture) model evolves slowly towards the (Gaussian) gravitational predictions but, even at $\sigma_8 = 1$, are still significantly larger, showing a characteristic minimum with a sharp increase in $S_J$ with increasing scales. This minimum, which is between 10 and $15\,h^{-1}\,\mathrm{Mpc}$, depending on the normalization, separates the regime where gravity starts dominating the evolution from the one in which the initial conditions are the dominant effect. In comparing this results with galaxy clustering observations, one has to take into account *biasing*, i.e. how galaxy fluctuations trace matter fluctuations. Although biasing could change the amplitudes, we show that the possible distortions to the shape of $S_J$ are typically small. In contrast to the non-Gaussian (texture) predictions, we find no significant minimum or rise in $S_J$ obtained from the APM Galaxy Survey.

## 1 INTRODUCTION

The large scale galaxy distribution can be used to study the origin and dynamics of cosmological fluctuations. The clustering of matter density fluctuations $\delta_R$, smoothed over scale $R$, is characterized here in terms of the reduced $J$-order moments $\overline{\xi}_J(R) \equiv \left\langle \delta_R^J \right\rangle_c$. Assuming that gravity is the dominant dynamical effect, the evolution of $\overline{\xi}_J(R,t)$ is completely fixed by the initial conditions $\overline{\xi}_J^0(R)$. For Gaussian inital conditions, i.e. $\overline{\xi}_J^0 = 0$ for $J > 2$, the leading order gravitational evolution for small $\overline{\xi}_2^0$, gives:

$$\overline{\xi}_J(R,t) = S_J^G(R)\,\overline{\xi}_2^{J-1}(R,t) \qquad (1)$$

where $\overline{\xi}_2$ follows the linear growth $\overline{\xi}_2(R,t) = a(t)^2 \overline{\xi}_2^0(R)$ (Peebles 1980). Time independent gravitational amplitudes, $S_J^G$, have been predicted using perturbation theory (PT) and tested against N-body simulations (Fry 1984; Juszkiewicz et al. 1993; Bernardeau 1994; Baugh et al. 1995). Gaztañaga & Baugh 1995 have shown how the values of $S_J^G$ change as a function of scale and time for different power spectrum, reproducing well the PT predictions on scales where $\overline{\xi}_2 \lesssim 1$.

Consider next the time evolution in the more general case of gravitational evolution from non-Gaussian initial conditions $\overline{\xi}_J^0 \neq 0$. The leading contribution gives:

$$\overline{\xi}_J(t) \simeq a^J\,\overline{\xi}_J^0 + a^{J+1}\,T_J\,\overline{\xi}_{J+1}^0 + a^{2(J-1)}\,S_J^G\,(\overline{\xi}_2^0)^{J-1} \qquad (2)$$

where $T_J$ are geometrical factors. In an homogeneous distribution $\overline{\xi}_J^0 \to 0$ in the limit $\overline{\xi}_2^0 \to 0$, so that we can write $\overline{\xi}_J^0 \to (\overline{\xi}_2^0)^\alpha$, with $\alpha = \alpha[J]$. When $\alpha > J-1$ the initial conditions are *forgotten*, as the leading order effect of the evolution is the hierarchical term in equation (1). When $J/2 < \alpha \leq J-1$ we have quasi-Gaussian but non-hierarchical initial conditions. Evolution in $\overline{\xi}_J$ have a dominant non-hierarchical term that grows as $a^J$, while the hierarchical term grows as $a^{2(J-1)}$ and may not become significant until $\overline{\xi}_2 \sim 1$. Note that, as pointed out by Fry & Scherrer 1994 that there is an additional non-Gaussian term that grows as $a^{J+1}$ which for the skewness $J=3$ contributes directly to $S_3 \equiv \overline{\xi}_3/\overline{\xi}_2^2$ so that $S_3 \neq S_3^G$ at all times. For $J > 3$ one expects a different scaling which could be used as a discriminating test (see Chodorowski & Bouchet 1996). If $\alpha < J/2$ there are strongly non-Gaussian initial conditions that dominate the evolution as far as $\overline{\xi}_2$ is small.

The question we want to address in this letter (see Silk & Juszkiewicz 1991) is to what extend gravity is able to erase the trace of the initial conditions in the transition case to the strong non-Gaussian regime $\alpha \simeq J/2$. This is important because large scale galaxy surveys can be used to find such a trace (Gaztañaga 1992; Bouchet et al. 1993; Gaztañaga 1995) Previous work by Weinberg & Cole 1992 provided interesting insights on different aspects of generic non-Gaussian models but did not address this question (other related works include Moscardini et al. 1993). Here we use a global texture model (Turok 1989) as a prototypical non-Gaussian model, which has recently been promoted as an interesting alternative for adiabatic inflationary structure formation (Cen et al. 1991).

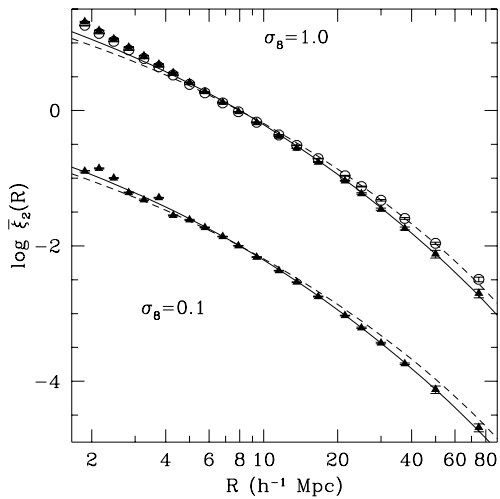

**Figure 1.** The variance $\overline{\xi}_2$ at two different epochs, $\sigma_8 = 0.1$ and $\sigma_8 = 1.0$, for the texture simulations (closed triangles). Open circles correspond to the Gaussian CDM $\Gamma = 0.5$ at $\sigma_8 = 1.0$. The dashed (continuous) lines show the linear PT predictions for the $\Gamma = 0.5$ ($\Gamma = 0.7$) CDM model at $\sigma_8 = 0.1$ and $\sigma_8 = 1.0$.

## 2 NON-GAUSSIAN TEXTURE MODEL

The non-Gaussian global texture model dynamics is governed by the evolution equation

$$\frac{\partial^2 \phi}{\partial \tau^2} + 2\frac{\dot{a}}{a}\frac{\partial \phi}{\partial \tau} - \nabla^2 \phi = -a^2 \frac{\partial V}{\partial \phi}, \qquad (3)$$

where $\phi$ is 4-dimensional scalar field, $\tau$ is conformal time and $V(\phi) = \frac{\lambda}{4}(\phi^2 - \phi_0^2)$ is the potential with the vacuum-energy value $\phi_0^2$. We have used a non-linear sigma model for the texture dynamics, evolving fields with the modified leap-frog integrator algorithm (Pen et al. 1994; Nagasawa et al. 1993) and calculating the density perturbations at each time step. The full texture dynamics is stopped at $\sigma_8 = 0.1$, when we map all density perturbations by $100^3$ particles to produce the initial conditions for a gravitational N-body simulation. The density fluctuations are then evolved by a $P^3M$-code (Efstathiou & Eastwood 1981; Efstathiou et al. 1988) until we reach $\sigma_8 = 1.0$. We have tested our simulations against the finite box size effects and our models are in good agreement with higher resolution simulations. The power spectrum $P(k)$ for the texture model turns out to be similar to the adiabatic inflationary CDM model characterized by $\Gamma = \Omega h$ (Bond & Efstathiou 1984). Here we use $\Omega = 1$ and $h = 0.5$. Our simulations show that texture models can be scaled as adiabatic models simply by $\Omega h$ for different matter densities within 10% accuracy. More details are given in Mähönen & Efstathiou 1995.

### 2.1 Clustering in the simulations

The volume averaged correlations $\overline{\xi}_J$ are estimated from moments of counts-in-cells as described in Baugh et al. 1995. We average the results over 3 simulations in each ensemble and use the dispersion between members to estimate the sampling errors. Different stages in the evolution of the sim-ulations are labeled by $\sigma_8$, the linear PT variance in spheres of radius $8\ h^{-1}$ Mpc, i.e. $\sigma_8^2 \equiv a^2 \overline{\xi}_2^0(8)$.

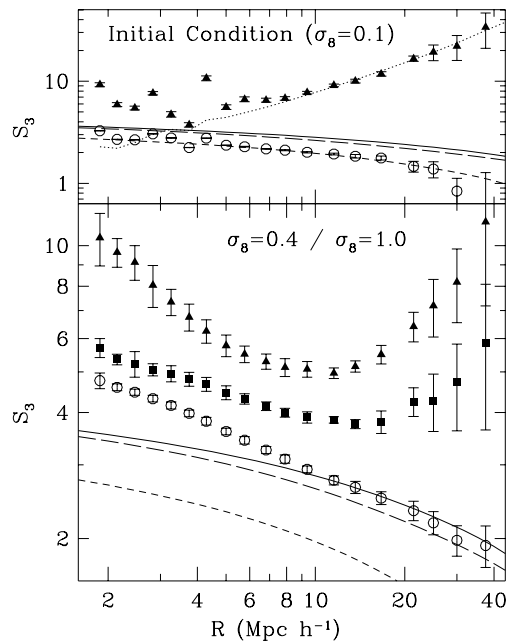

**Figure 2.** The skewness $S_3 = \overline{\xi}_3/\overline{\xi}_2^2$ in the Gaussian $\Gamma = 0.5$ CDM model (open circles) compared with the Non-Gaussian model (filled symbols). The top panel shows the initial conditions, $\sigma_8 = 0.1$, while the bottom shows time $\sigma_8 = 0.4$ (triangles) and $\sigma_8 = 1.0$ (squares). Lines show the Zeldovich approximation (short-dashed) and the perturbation theory predictions for the $\Gamma = 0.5$ (continuous) and $\Gamma = 0.7$ (long-dashed) models.

Figure 1 shows the variance in the texture model (filled triangles) at two different epochs, $\sigma_8 = 0.1$ and $\sigma_8 = 1.0$. Note that the initial results match roughly the linear variance in the CDM $\Gamma = 0.5$ model (dashed-line), although in detail they are closer to the $\Gamma = 0.7$ shape (continuous line). The evolved non-linear variance at $\sigma_8 = 1.0$ is close to the corresponding non-linear variance in the Gaussian $\Gamma = 0.5$ simulations (open circles). Note that the deviations from the linear growth (represented by the upper lines) are similar in both cases, indicating that the initial non-Gaussianities have only a small effect in the non-linear growth.

The top panel in Figure 2 shows the initial values of the normalized skewness: $S_3 \equiv \overline{\xi}_3/\overline{\xi}_2^2$ in both models. The Gaussian model (open circles) matches well with the Zeldovich approximation, as expected (see Baugh et al. 1995). The non-Gaussian model shows a characteristic increase of $S_3$ with scale. At large scales, $R \gtrsim 10\ h^{-1}$ Mpc, a fit of the form $\overline{\xi}_3 = A\overline{\xi}_2^\alpha$ yields $A \simeq 1$ and $\alpha \simeq 3/2 + 0.1$ (dotted line in Figure 2). This is close to the strongly non-Gaussian transition mentioned in the introduction. The lower panel in Figure 2 shows the evolution of $S_3$ for Gaussian models (open circles) at $\sigma_8 = 1$ in comparison with the non-Gaussian models at $\sigma_8 = 0.4$ (triangles) and $\sigma_8 = 1$ (squares). The Gaussian models reproduce quite well the PT predictions at large scales. As the texture simulations evolve, the shape and amplitude of $S_3$ in the non-Gaussian model slowly approaches the one in the Gaussian models at small scales. At larger scales there is a change in the slope of $S_J$, showing a characteristic minimum which separates the regime where gravity starts dominating the evolution from the one in which the initial conditions are still the dominant effect, i.e. equation (2). The scale at which the minimum occurs $R_{min}$ is a function of time (i.e. $\sigma_8$) moving from $R_{min} \simeq 10\ h^{-1}$ Mpc at $\sigma_8 = 0.4$ to $R_{min} \simeq 15\ h^{-1}$ Mpc at $\sigma_8 = 1$.

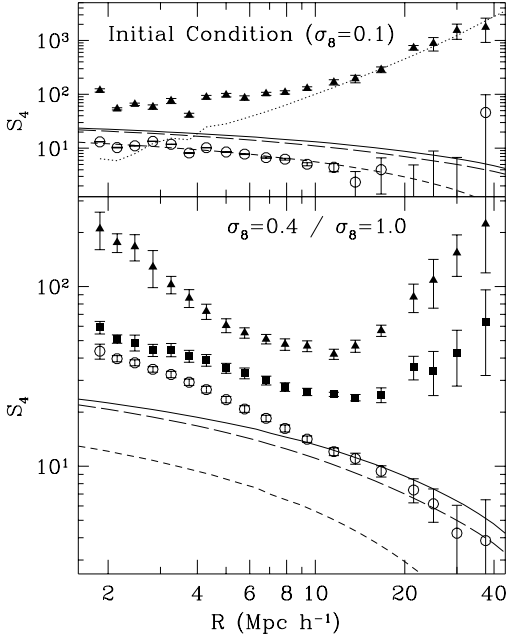

**Figure 3.** The kurtosis $S_4 = \overline{\xi}_4/\overline{\xi}_2^3$ as in Figure 2.

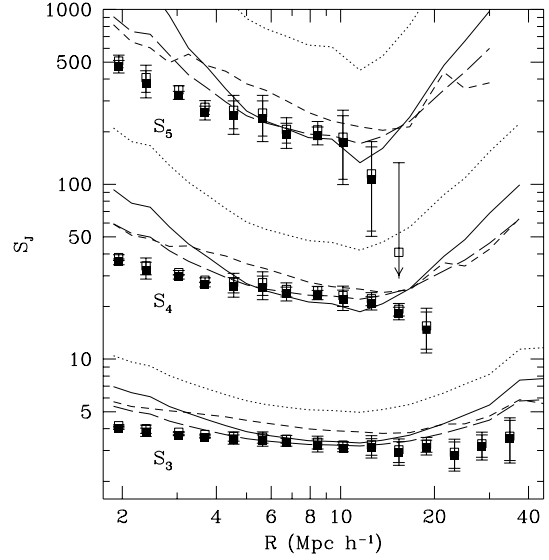

**Figure 4.** Amplitudes $S_J$ for $J = 3-5$ in the APM galaxies (symbols) compared with the matter amplitudes in the Non-Gaussian texture models normalized to $\sigma_8 = 1$ (short-dashed line), to $\sigma_8 = 0.4$ (dotted line) and scaled with a linear bias $b = 1.5$ (continuous line) and non-linear bias $b = 2.5$ (long-dashed line).

A similar trend follows for higher order amplitudes $S_J$ (see Figure 3 and Figure 4). We find that the initial conditions follow: $\overline{\xi}_J = \overline{\xi}_2^{J/2+\epsilon}$, with $\epsilon \simeq 0.1$ (dotted line in Figure 3), again close to the transition to strongly non-Gaussian initial conditions. This tendency agrees well with the predictions for $J = 3 - 4$ by Turok & Spergel 1991. The time evolution, at intermediate scales $R \simeq 10\ h^{-1}$ Mpc, can be approximated by:

$$\overline{\xi}_J(z) = (1+z)^{0.2(J-1)} \overline{\xi}_J(z=0) \qquad (4)$$

between $z = 0$ ($\sigma_8 = 1.0$) and $z = 1.5$ ($\sigma_8 = 0.4$).

## 3 COMPARISON WITH THE APM

In Figure 4 we show the values of $S_3$, $S_4$ and $S_5$ estimated from the angular APM Galaxy Survey (Gaztañaga 1994), assuming no evolution in $S_J$ (closed symbols) or the texture evolution (opened symbols) given by equation (4). The model for the evolution leads only to small differences, since the mean redshift in the APM is only $\bar{z} \simeq 0.15$. These three-dimensional estimates result from using a simple scaling law to model the projection effects. Although there are some potential problems with this modeling Bernardeau 1995, we believe that these results are accurate (see Gaztañaga 1995;

Baugh & Gaztañaga 1996). Depending on the way counts-in-cells are estimated the mean angular amplitudes could increase or decrease rapidly with scale at the largest scales, i.e. $l \gtrsim 10^\circ$ ($\gtrsim 40\ h^{-1}$ Mpc). These diverging estimates are not reliable as sampling effects from the finite APM volume dominate the statistics at these larger scales (Gaztañaga 1994; Baugh & Gaztañaga 1996).

Large scale galaxy fluctuations, $\delta_g$, might be biased tracers of the underlying matter fluctuations, $\delta$. To account for this possible bias and uncertainties in the normalization of the texture model, we consider different outputs and scale them with different biasing prescriptions. In Figure 4 we show $S_J$ in the the texture model for two different outputs, together with the values at $\sigma_8 = 0.4$ normalized with a linear biasing relation, $\delta_g = b\ \delta$ which produces $S_{J,g} = S_J/b^{J-2}$. We have chosen $b = 1.5$ as the optimal value to match the APM amplitudes $S_{J,g}$ around $8\ h^{-1}$ Mpc. Note that for $\sigma_8 = 0.4$ the linear bias requires $b = 2.5$ if we want to fit the APM variance at $8\ h^{-1}$ Mpc, but this value of $b$ produces a poor matching for $S_{J,g}$. We introduce more biasing parameters with a non-linear transformation:

$$\delta_g = f[\delta] \simeq b\ [\ \delta + \frac{c_2}{2!}\delta^2 + \frac{c_3}{3!}\delta^3 + \frac{c_4}{4!}\delta^4 + ...\ ] \qquad (5)$$

which for small variances, $\overline{\xi}_2 < 1$, still gives a linear relation for $\overline{\xi}_2$ but changes the final amplitudes to $S_{g,J}$ given by equation (10) in Fry & Gaztañaga 1993, e.g. $S_{3,g} = (S_3 + 3c_2)/b$. For the texture amplitudes at $\sigma_8 = 0.4$ we have to fix $b = 2.5$, $c_2 \simeq 1$, $c_3 \simeq 6$ and $c_4 \simeq -180$ to match $\overline{\xi}_2$, $S_3$, $S_4$ and $S_5$ at $8\ h^{-1}$ Mpc in the APM. The resulting shapes are shown as long-dashed lines in Figure 4. For scales up to $40\ h^{-1}$ Mpc in $S_3$, or up to $20\ h^{-1}$ Mpc in $S_4$ we find no significant minimum or rise within the errors in the APM, in contrast to the unbiased or biased texture predictions.

## 4  CONCLUSION

In Gaussian models with different initial power spectrum there is an excellent agreement for $S_J$ between PT and N-body simulations on scales where the variance is approximately linear Juszkiewicz et al. 1993; Bernardeau 1994; Gaztañaga & Baugh 1995. The values of $S_J$ do not evolve much with time. In contrast, we have seen here that the non-Gaussian (texture) model show a strong evolution from large initial values of $S_J$ towards the values found in the corresponding Gaussian model (with similar initial power spectrum), although even at $\sigma_8 = 1$ there are important differences. The non-Gaussian models have a characteristic minimum that separates the regime where gravity starts dominating the evolution from the one in which the initial conditions are still the dominant effect. The scale at which the minimum occurs is a function of time (and therefore of the normalization) moving from $R_{min} \simeq 10\,h^{-1}$ Mpc at $\sigma_8 = 0.4$ to $R_{min} \simeq 15\,h^{-1}$ Mpc at $\sigma_8 = 1$.

Our conclusions for the texture model seem different from those in Nagasawa et al. 1993, but a clear comparison can not be made as their results apply only to smaller scales and do not include the late-time gravitational dynamic that is simulated here with a full N-body simulation.

A direct comparison of the mass amplitudes $S_J$ with amplitudes in the APM, is not very useful, as the texture model does not even reproduce the second order statistics. The variance $\overline{\xi}_2$ has less power at large scales than the APM galaxy distribution, which has a shape $\Gamma \simeq 0.2$ (see Maddox et al. 1990; Gaztañaga 1995). A biasing between matter and galaxy fluctuations might account for part of this difference at small scales, but at large scales biasing is unlikely to introduce distortions on the shape of $\overline{\xi}_2$. A texture model with more power on large scales (e.g. $\Omega < 1$), but with similar initial conditions, would produce similar results for $S_J$, but with slightly larger gravitational amplitudes, $S_J^G$.

Biasing could also affect the final, directly observable, (galaxy) amplitudes $S_{J,g}$ (e.g. Fry & Gaztañaga 1993) and could even introduce scale dependence distortions for large biasing parameters (see Gaztañaga & Frieman 1994). Nevertheless, we have shown that these distortions are typically small in comparison with the pronounced non-Gaussian features that we find. Thus, one would still expect $S_J$ to show a minimum or a steep rise around $R \simeq 15\,h^{-1}$ Mpc. For scales up $40\,h^{-1}$ Mpc in $S_3$ or $20\,h^{-1}$ Mpc in $S_4$, we do not find this characteristic minimum or steep rise in the APM fluctuations, in contrast to the predictions in the non-Gaussian (texture) model.

### Acknowledgements

We would like to thank George Efstathiou for his help and advice to produce the N-body simulations used in this letter, and also Neil Turok for providing a copy of his original non-linear sigma model texture code. EG acknowledges support from DGICYT (Spain), project PB93-0035 and CIRIT (Generalitat de Catalunya), grant GR94-8001. PM acknowledges financial support from the Academy of Finland. We thank Finnish Center for Scientific Computing for supercomputer facilities.